\documentclass[12pt,english]{article} \usepackage{graphicx}
\usepackage{latexsym} \usepackage{amssymb} 
\usepackage{amsmath,amsfonts}
\newcommand{\half}{\frac{1}{2}}

\newcommand{\labell}[1]{\label{#1}}  
\newcommand{\reef}[1]{(\ref{#1})}  
  
\newcommand{\href}[2]{#2}

\DeclareSymbolFont{AMSb}{U}{msb}{m}{n}
\DeclareMathSymbol{\IN}{\mathbin}{AMSb}{"4E}
\DeclareMathSymbol{\IZ}{\mathbin}{AMSb}{"5A}
\DeclareMathSymbol{\IR}{\mathbin}{AMSb}{"52}
\DeclareMathSymbol{\Q}{\mathbin}{AMSb}{"51}
\DeclareMathSymbol{\II}{\mathbin}{AMSb}{"49}
\DeclareMathSymbol{\IC}{\mathbin}{AMSb}{"43}
\DeclareMathSymbol{\IP}{\mathbin}{AMSb}{"50}
\DeclareMathSymbol{\IH}{\mathbin}{AMSb}{"48}
\DeclareMathSymbol\IA{\mathalpha}{AMSb}{"41}
\DeclareMathSymbol\IS{\mathalpha}{AMSb}{"53}

\def\Q{{\cal Q}}

\begin{document}  
  
\bigskip  
\hskip 4.7in\vbox{\baselineskip12pt  
\hbox{hep-th/0303255}}  
  
\bigskip  
\bigskip  
\bigskip

\begin{center}
  {\Large \bf A Note on D--brane --- Anti--D--brane
    Interactions}\\\bigskip{\Large \bf in Plane Wave Backgrounds}
  \end{center}
\bigskip  
\bigskip  
\bigskip  
\bigskip 
  
\centerline{\bf Clifford V. Johnson, Harald G. Svendsen}

\bigskip  
\bigskip 
\bigskip

\centerline{\it Centre  
for Particle Theory}
  \centerline{\it Department of Mathematical Sciences}  
\centerline{\it University of  
Durham}
\centerline{\it Durham, DH1 3LE, U.K.}  

\centerline{$\phantom{and}$}  

\bigskip

\centerline{\small \tt  
  c.v.johnson@durham.ac.uk, h.g.svendsen@durham.ac.uk}  
  
\bigskip  
\bigskip  
\bigskip  

  
\begin{abstract}  
  \vskip 4pt We study aspects of the interaction between a D--brane
  and an anti-D--brane in the maximally supersymmetric plane wave
  background of type~IIB superstring theory, which is equipped with a
  mass parameter $\mu$.  An early such study in flat spacetime
  ($\mu=0$) served to sharpen intuition about D--brane interactions,
  showing in particular the key role of the ``stringy halo'' that
  surrounds a D--brane. The halo marks the edge of the region within
  which tachyon condensation occurs, opening a gateway to new
  non--trivial vacua of the theory.  It seems pertinent to study the
  fate of the halo for non--zero $\mu$.  We focus on the simplest
  cases of a Lorentzian brane with $p=1$ and an Euclidean brane with
  $p=-1$, the D--instanton. For the Lorentzian brane, we observe that
  the halo is unaffected by the presence of non--zero $\mu$. This most
  likely extends to other (Lorentzian) $p$.  For the Euclidean brane,
  we find that the halo is affected by non--zero~$\mu$. As this is
  related to subtleties in defining the exchange amplitude between
  Euclidean branes in the open string sector, we expect this to extend
  to all Euclidean branes in this background.
\end{abstract}  
\newpage  
\baselineskip=18pt  
\setcounter{footnote}{0}  
  
  
\section{Introduction} 

A D--brane and its ``anti--particle'', an anti--D--brane, upon
approaching each other, will annihilate. The generic product of this
annihilation process is expected to be a state of closed strings,
which carry no net R--R charge.  This expectation is supported by
field theory intuition and knowledge of which objects are the carriers
of the available conserved charges in perturbative string theory. From
experience with field theory one expects to be able to see the
beginnings of the process of annihilation {\it via} the opening up of
new decay channels at coincidence. These can be seen by studying the
amplitude for exchange of quanta between the two branes, which gives a
potential.  At small separations, the behaviour of the interaction
potential can signal new physics. Basically, a divergence in the
amplitude as the objects are brought together can signal the opening
up of a new channel (or new channels) not included in the computation
of the amplitude away from the divergent regime.

In field theory, for  a separation $X$ of the two objects, the
divergence follows simply from the fact that the amplitude for
exchange is controlled by the position space propagator $\Delta(X)$
which (for more than two transverse directions) is divergent at $X=0$.
This is where the new channels can open up, which can include the
processes for complete annihilation into a new sector, if permitted by
the symmetries of the theory.

For D--branes in superstring theory, such a divergence does indeed
show up, but there is an important new feature\cite{Banks:1995ch}. The
divergence occurs when the D--branes are finitely separated, by an
amount set by $X_H^2=2\pi^2 \alpha^\prime$, where $\alpha^\prime$ is
the characteristic length scale set by a fundamental string's tension.
This is interpreted as the fact that in addition to the many special
features of D--branes, they have a ``stringy halo'' originating in the
fact that the bulk of the open strings which (by definition) end on
them can reach out in the transverse directions, forming a region of
potential activity of size set by $X_H$. This halo means that the
D--branes can interact with each other before zero separation, as
there is an enhancement of the physics of interaction by new light
states formed by the entanglement of the halos, and the crossover into
the annihilation channel begins before the branes are coincident.

Recall that the amplitude of exchange can be thought of using two
equivalent pictures: Either as tree level exchange of closed string
quanta between the branes, or (after a modular transformation) as the
one--loop vacuum diagram for open strings stretched between the two
D--branes. In the open string description, at separation $X_H$, the
lightest open string becomes massless, and for any closer separation
it becomes tachyonic, signalling that the entire vacuum configuration
is unstable and wishes to roll to another vacuum. It is this tachyon
which produces the divergence in the amplitude, converting a
decaying exponential into a growing one, spoiling the convergence of
the amplitude in the infra--red (IR) region.

The D--branes annihilate {\it via} conversion to closed strings in the
generic situation, but the tachyon picture can be exploited in a
beautiful way to produce more
structure\cite{Sen:1998rg,Sen:1998ii,Sen:1998sm,Sen:1998tt}. For the
$G=U(N)\times U(N)$ gauge theory on the $(p+1)$--dimensional
world--volume on $N$ D$p$--branes and $N$ anti--D$p$--branes, the tachyon
field, transforming as the $({\mathbf N},{\bar {\mathbf N}})$, can be put into a
configuration endowed with non--trivial topological charge, and the
tachyon potential need not yield a runaway to a sector containing only
closed strings.  Having such topological vacuum solutions in the
tachyon sector allows for the possibility of a stable remnant ---
interpreted as a D--brane of lower dimension--- of the annihilation
process after the debris that is the closed string products has
cleared. It turns out that the spectrum of hypermultiplets in the
$U(N)\times U(N)$ world--volume theory supplies a set of variables
which is isomorphic to those needed to perform a K--theoretic analysis
of the topology of $G$--vector bundles over the world--volume, and so
the classification of all D--branes which can appear on a spacetime is
apparently elegantly and economically by using the results of the
appropriate K--theory of the spacetime which the D$p$--branes and
anti--D$p$--branes
fill\cite{Minasian:1997mm,Witten:1998cd,Horava:1998jy}. The case of
$p=9$ for Minkowski spacetime yields the entire classification of
D--branes in the most familiar symmetric vacuum of type~IIB superstring theory.

This is all well understood for the case of flat ten dimensional
spacetime. So when one encounters another background  which enjoys the
same maximal supersymmetry as flat spacetime --- a  plane wave with
R--R flux\cite{Blau:2001ne}:
\begin{eqnarray}
 ds^2 &=& 2dx^+dx^- - \mu^2 x^2(dx^+)^2 
  + \sum_{i=1}^{4} dx^i dx^i + \sum_{i=5}^{8} dx^i dx^i\ ,\nonumber \\
&&F_{+1234}=F_{+5678}=2\mu\ , \qquad x^2=\sum_{i=1}^8 x^i x^i\ ,
  \qquad x^\pm = \frac{1}{\sqrt{2}}(x^9\pm x^0)\ ,
  \label{eq:planewave}
\end{eqnarray}
which also yields an exactly solvable string
model\cite{Metsaev:2001bj} (in light--cone gauge defined by relating
worldsheet time $\tau$ to $x^+$ {\it via} $x^+ = 2\pi\alpha' p^+\tau$,
where $p^+$ is the $+$ component of spacetime momentum):
\begin{equation}
  \mathcal{L} = \frac{1}{4\pi\alpha'}
  (\partial_+ x^i\partial_- x^i - M^2x^2)
  +\frac{i}{2\pi\alpha'}
  (S^a\partial_ + S^a + \tilde{S}^a\partial_-\tilde{S}^a 
  - 2MS^a\Pi_{ab}\tilde{S}^a)\ ,
\labell{exactlysolvable}
\end{equation}
with a mass parameter $M=2\pi\alpha' p^+\mu$ --- it is inevitable that
questions about the key lessons which were learned about D--branes
will spring to mind\footnote{There has been a number of papers
  studying D--branes in plane wave and pp--wave backgrounds. Some of
  them are
  refs.\cite{Dabholkar:2002zc}--\cite{Panigrahi:2003rh}.}.
Is the picture of D--branes as Dirichlet open string boundary
conditions as powerful in this context as it has been in flat spacetime?
In particular, do the dynamics hidden within a halo's breadth of the
branes bear any similarity to the flat spacetime case?  Are all D--branes
classified by K--theory, now of the new background?

In this paper we note that the properties of the halo ---the fact that
it exists, and also its location and size--- are unaffected by
non--zero $\mu$ for all branes that have a Lorentzian definition, {\it
  i.e.}, are at a definite position in space, but not time. So this
particular (and important) property of D--branes in this non--trivial
R--R background is very much like that in flat space. This bodes well
for an attempt to classify such D--branes in this background using
tachyon condensation and K--theory. Howver, for branes with a
Euclidean definition, such as the $p=-1$ brane, we find that the halo
---or at least its analogue in this context--- is deformed by non--zero
$\mu$.

\section{The Interaction}

It is convenient\cite{Skenderis:2002vf,Gaberdiel:2002hh} to label
D--branes in the plane wave background given in
equation~\reef{eq:planewave} as $(r,s)$, if they are Euclidean, where
$r$ denotes the spatial extent in directions $i=1,2,3,4$ and $s$
denotes the spatial extent in directions $i=5,6,7,8$. A D$p$--brane
would then have $r+s=p+1$. If the D--branes are Lorentzian, then their
worldvolume extends in the $x^+$ and $x^-$ direction, and the notation
is $(+,-,r,s)$. In that case, a D$p$--brane has $r+s=p-1$.

\begin{figure}[htbp]
  \centering
  \includegraphics[height=6cm]{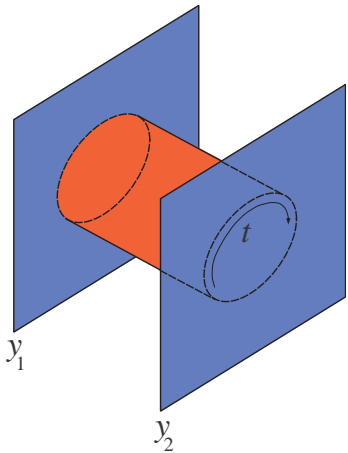}
  \caption{\small Cylinder diagram for computing the amplitude of interaction
  between two branes. The parameter $t$ is open string propagation
  time, and is the modulus of the cylinder.}
  \label{cylinder_exchange}
\end{figure}

The string theory diagram of interest is a cylinder, representing
either the tree level exchange of closed string quanta between two
D--branes, or the one--loop vacuum process involving the circulation
of open strings with ends on either D--brane. 
See figure~\ref{cylinder_exchange}.

We will focus on the results for the simplest branes in the Euclidean
and Lorentzian classes. These are the D$(-1)$--branes (or
$(0,0)$--branes), and the D1--branes (or $(+,-,0,0)$--branes), discussed
in ref.\cite{Gaberdiel:2002hh}. The former requires the time
direction, in which the branes are also pointlike, to be Euclidean.

The results are reasonably simple for these cases, compared to other
$(r,s)$ with $r\neq s\neq 0$, and it would be interesting to explore those
other cases in detail.  We expect that the key
observations made in this paper for these $r=0=s$ cases will be quite
generic, although there may be additional features to be deduced from
studying other cases in detail.

\subsection{The Amplitude and Potential}

We consider a D$p$--brane and its antiparticle for $p=\pm1$. If
$p=-1$, it is an instanton, (a $(0,0)$--brane) and we consider it to
be pointlike in Euclidean time.  If $p=+1$ it is a string, (a
$(+,-,0,0)$--brane) and the theory is Lorentzian. 

So we place a D$p$--brane at position $y_1^i$ in the $x^i$ directions
($i=1,\ldots,8$), and a $\overline{{\rm D}p}$--brane (anti--brane) at
position $y_2^i$, with a separation $X^\pm$ in the $x^\pm$ directions
if $p=-1$. The cylinder amplitude $A$ is\cite{Gaberdiel:2002hh}:
\begin{equation}
  A = \int_0^\infty \frac{dt}{2t} t^{-\left(\frac{p+1}{2}\right)}e^{it\frac{X^+ X^-}{2\pi\alpha'}}
 \hat{h}_0({t};y_1,y_2) \frac{\hat{g}_4^{({m})}({t})^4}
  {f_1^{({m})}({t})^8}\ .
\labell{amplitudeform}
\end{equation}
where we have performed a double Wick rotation: $\tau\to i t$, $x^+\to
i x^+$.
For $p=+1$, the factor $\exp(itX^+X^-/2\pi\alpha^\prime )$ is not
present. For higher $p$, (which we will not be considering here) there
are no additional powers of $t$ in the integrand. These are normally
due to integration over continuous zero modes in the flat spacetime case.
The plane wave background has no such modes for the directions $x^i$,
(the zero modes are instead themselves harmonic
oscillators\cite{Amati:1989sa,Horowitz:1990bv,Horowitz:1990sr,Jofre:1994hd})
and so no such $t^{-1}$ factors beyond those appearing here are present.
See below equation~\reef{marker} for some further discussion of how to
read this expression.

In the above, 
we have the functions:
\begin{align}
  f_1^{(m)}(t) = & q^{-\Delta_m}(1-q^m)^\half \prod_{n=1}^{\infty}
  ( 1-q^{\omega_n} ),
  \nonumber \\
  \hat{h}_0({t};y_1,y_2) = & \exp\Bigl(
  -\frac{{m}{t}}{2\alpha'\sinh(\pi{m})}
  [\cosh(\pi{m})(y_1^2+y_2^2) - 2 y_1\cdot y_2] \Bigr),
  \nonumber \\
  \hat{g}_4^{({m})}({t}) =
  &{q}^{-\hat{\Delta}_{({m})}} 
  \prod_{l\in\mathcal{P}_+}\Bigl(1-{q}^{|{\omega}_l|}\Bigr)^\half
  \prod_{l\in\mathcal{P}_-}\Bigl(1-{q}^{|{\omega}_l|}\Bigr)^\half,
  \nonumber \\
 \Delta_m = & - \frac{1}{(2\pi)^2}\sum_{p=1}^\infty \int_0^\infty ds
  \,\, e^{-p^2 s - \frac{\pi^2 m^2}{s}} \ ,\nonumber \\  
\hat{\Delta}_{{m}} = & - \frac{1}{(2\pi)^2}\sum_{p=1}^\infty 
  (-1)^p\sum_{r=0}^\infty c_r^p {m} 
  \frac{\partial^r}{(\partial {m}^2)^r} \frac{1}{{m}}
  \int_0^\infty ds \left(\frac{-s}{\pi^2}\right)^r
  e^{-p^2s-\frac{\pi^2{m}^2}{s}}\ ,
\end{align}
and the parameter $q$ and the deformed harmonic oscillator frequencies
are defined as:
\begin{equation}
  q = e^{-2\pi t}\ ,\qquad \omega_n = \mathrm{sign}(n)\sqrt{n^2+m^2} \ .
\end{equation}
Note that $\Delta_m$ and $\hat{\Delta}_{{m}}$ are zero--point energies
which arise naturally in the closed and open string sectors,
respectively.  The coefficients $c_r^p$ in $\hat{\Delta}_{{m}}$ are
the coefficients of a specific Taylor expansion:
\begin{equation}
  \Bigl(\frac{x+1}{x-1}\Bigr)^p + \Bigl(\frac{x-1}{x+1}\Bigr)^p =
  \sum_{r=0}^\infty c_r^p x^{2r}\ . 
\end{equation}
The sets $\mathcal{P}_-$ and $\mathcal{P}_+$ are given as solutions of
the equations 
\begin{align}
  l\in \mathcal{P}_- :\quad 
  \frac{l-im}{l+im}+e^{2\pi i l}=0\ ,
  \qquad
  l\in \mathcal{P}+ :\quad 
  \frac{l+im}{l-im}+e^{2\pi i l}=0\ .
\label{marker}
\end{align}
The details of the derivation of these amplitudes can be found in
ref.\cite{Gaberdiel:2002hh}. We will not need them all here, and refer
the reader there for more information.  Some comments are in order
however. For the case $p=1$, the computation was done directly in
terms of the open string channel, with open string light cone gauge
$x^+=2\pi\alpha^\prime p^+ \tau$, so we have
\begin{equation}
  \quad t = \frac{X^+}{2\pi\alpha' p^+}\ ,
  \quad {m}=2\pi\alpha^\prime p^+\mu\ .
\label{masstwo}
\end{equation}

For the case $p=-1$, however, things are more subtle. A Dirichlet
condition is needed in the time direction, but this is incompatible
with the standard light--cone gauge choice. The amplitude is defined
by appealing to the open--closed string duality instead. The amplitude
is defined in the closed string sector by propagating for a distance
$X^+$ between two boundary states. The propagation time is 
\begin{displaymath}
  {\tilde t}=\frac{X^+}{2\pi \alpha^\prime p^+}\ ,
\end{displaymath}
since light--cone gauge in the closed string sector is
$x^+=2\pi\alpha^\prime p^+ \tau$, with mass parameter
\begin{equation}
M=2\pi\alpha^\prime p^+\mu\ .
 \label{closedmass}
\end{equation}
Open--closed string duality is then invoked to define the amplitude
given in equation~\reef{amplitudeform}, where
modular transformation gives the expression above, with
\begin{equation}
t=1/{\tilde t} = \frac{2\pi\alpha' p^+}{X^+}\ ,
  \quad \mbox{\rm and}\quad{m}=\mu X^+   =Mt^{-1}\ .
\labell{massone}
\end{equation}
This will be very important later.

\subsection{Divergences, Tachyons, and the Halo}

What is important for our discussion is the structure of the full
amplitude for the cylinder diagram, given above in
equation~\reef{amplitudeform} as an integral over the modulus $t$.  
It can be written as:
\begin{equation}
  A =
 \int_0^\infty \frac{dt}{2t} t^{-\left(\frac{p+1}{2}\right)}\exp\Bigl\{-2\pi t
Z(m,y_1,y_2)\Bigl\} G(t),
\end{equation} 
where the the exponent $Z(m,y_1,y_2))$ is defined as (delete the
$X^+X^-$ term to get the D1--brane result):
\begin{equation}
\begin{split}
  Z(m,y_1,y_2) = 
\frac{m\pi}{4\pi^2\alpha'\sinh(m\pi)}
         \left[ \cosh(m\pi)(y_1^2+y_2^2)-2y_1\cdot y_2\right] 
    - 4(\hat{\Delta}_m-2\Delta_m)
    -i\frac{X^+X^-}{4\pi^2\alpha'}.
\end{split}
\end{equation} and the
function $G(t)$ is defined as:
\begin{equation}
\begin{split}
  G(t) 
= \frac{\prod_{l\in\mathcal{P}_+}(1-q^{|\omega_l|})^2 
    \prod_{l\in\mathcal{P}_-}(1-q^{|\omega_l|})^2}
  {(1-q^m)^4 \prod_{n=1}^{\infty}(1-q^{\omega_n})^8}
= \prod_{n=1}^{\infty}(1-q^{\omega_n})^{-8}
  \prod_{l\in\mathcal{P}_+, l>0} (1-q^{\omega_l})^4
  \prod_{l\in\mathcal{P}_-, l>0} (1-q^{\omega_l})^4  
\end{split}
\end{equation}
For our discussion, the only important fact about the function $G(t)$
is that its behaviour at large and small $t$ is such that generically,
the amplitude is convergent.  That $A$ is finite as $t\to 0$ follows
from the fact that small $t$ is the closed string IR limit, where this
amplitude should reproduce simple low energy field theory results for
massless exchange at tree level. The $t\to\infty$ limit is also well
behaved generically, since this is the open string IR limit,
which is fine --- away from special circumstances which will not show
up in the oscillator contributions since their energies are higher
than the lowest lying states. In fact, it is clear that $G(t)\to 1$ as
$t\to\infty$, and so whether $A$ is finite as $t\to\infty$ depends on
the sign of the exponent $Z$, which 
controls those lowest lying states.

 The divergence
for negative $Z$  is related
to the lowest lying states becoming tachyonic at this point, as is
most familiar in the RNS formulation in the flat spacetime background.
Then the worldsheet Hamiltonian is given as $H=L_0=\alpha' p^2 + N +
a_{R(NS)}$, where the constant $a_{R(NS)}$ is the zero point energy
and~$N$ is the total number operator.  The z.p.e. is $a_R=0$ in the
Ramond sector, and $a_{NS}=-\half$ in NS sector.

For strings stretched between two D--branes, we have $p^m =
{x^m}/{2\pi\alpha'}$ for transverse (to the branes) directions
$x^m$.  So, splitting transverse (labelled $m$) and parallel (labelled
$i$) directions we can write
\begin{equation}
  L_0 = \alpha' p^i p_i + N + \frac{z^2}{4\pi^2\alpha'} + a_{R(NS)}\ .
\end{equation}
This gives a mass spectrum
\begin{equation}
  M^2 = -p^ip_i = \frac{1}{\alpha'}
\left(N +a_{R(NS)} + \frac{z^2}{4\pi^2\alpha'}\right)\ .
\end{equation}
The NS ground state ($N=0$, $a_{NS}=-\half$) has mass squared
\begin{equation}
M_0^2 = \frac{1}{2\alpha'}\left(\frac{z^2}{2\pi^2\alpha'}-1\right)\ .
\end{equation}
This is a tachyon if $z^2 < 2\pi^2\alpha'$.

In the usual case this ground state is eliminated by the GSO
projection $P=\frac{1+(-1)^F}{2}$ in superstring theory. 
When we consider a brane--anti--brane system, we are effectively
reversing the GSO projection in the partition function, giving
$P=\frac{1-(-1)^F}{2}$, since anti--branes come with a minus sign.  This
means that the NS ground state ($N=0$) will now survive, and the
possible tachyon above is present in the spectrum. So for $z^2<
2\pi^2\alpha'$ there is a tachyon, and so there is a 1--1
correspondence between the tachyon's appearance and divergence of the
integral. (For the case when all of the directions are transverse, as
is the case for D--instantons, the tachyon interpretation follows from
continuation and T--duality.)

Let us write everything in terms $z^i$, the separation between the
branes in the eight directions~$x^i$, defined by $y_2^i=y_1^i+z^i$.
The expression for $Z$ then  becomes
\begin{equation}
  Z(m,y_1,z) = \frac{1}{4\pi^2\alpha'}\frac{m\pi}{\tanh(m\pi)} \Bigl[
      (z+ a)^2 -i\frac{\tanh(m\pi)}{m\pi}X^+X^-- b^2 \Bigr],
\labell{complicatedZ}
\end{equation}
where we have defined:
\begin{align}
  a =  \frac{\cosh(m\pi)-1}{\cosh(m\pi)}y_1\ ,\qquad
  b =  \tanh(m\pi)\sqrt{y_*^2-y_1^2}\ ,\qquad
  y_*^2 =  \frac{16\pi^2\alpha' (\hat{\Delta}_m - 2\Delta_m)}
  {m\pi \tanh(m\pi)}\ .
\labell{crucial}
\end{align}

For the Lorentzian $p=1$ case, these parameters simplify further in
the $t\to\infty$ limit of interest. Since for fixed $X^+$ the large
$t$ region corresponds to small $p^+$ (this follows from
equation~\reef{masstwo}, or on general grounds from the operator
definition of the amplitude), we see that  $m\to0$ in all
of these expressions, and so we obtain:
\begin{equation}
  Z\longrightarrow \frac{1}{4\pi^2\alpha'}(z^2-2\pi^2\alpha') \ ,
\end{equation}
This is in fact the same expression one would obtain from the
equivalent flat space computation, which simply has $m=0$ throughout,
and so we recover the well known\cite{Banks:1995ch} divergence at
separation given by $X_H^2=2\pi^2\alpha'$. In fact, the result ought
to be present for all Lorentzian branes, as the relevant amplitude can
be defined directly in the open string light cone gauge. Intuitively,
we are looking for a result in the open string IR limit $t\to\infty$,
which (from equation~\reef{masstwo}) corresponds to $p^+\to0$. But the
parameter upon which any new physics can depend is
$m=2\pi\alpha^\prime p^+\mu$, which vanishes in the limit. So there is
no new physics.

For the Euclidean $p=-1$ case, the situation is very different. Now,
for a given separation $X^+$, the $t\to\infty$ limit corresponds (due
to equation~\reef{massone}) to $p^+\to\infty$ (this is the {\it closed
  string} momentum) and so things get quite reversed.  In fact, the
natural mass parameter seen by the open string physics is $m=\mu X^+$.
In fact, there is quite a complicated dependence on $m$, as is evident
from the equation~\reef{complicatedZ}.  Looking (without loss of
generality, since the spacetime is homogeneous\footnote{We thank Simon
  Ross for reminding us of the significance of this for
  D--instantons.}) at the case where we put one brane at the origin in
the transverse directions, and so $y_1^i=0$ and $z^i=y_2^i$, then the
vanishing of $Z$ can be written:
\begin{equation}
z^2-i\frac{\tanh(\pi\mu X^+)}{\pi\mu X^+} X^+X^-=2\pi^2 \alpha^\prime {\cal D}(\mu X^+)\frac{\tanh(\pi\mu X^+)}{\pi\mu X^+}\ ,
  \label{vanisihngone}
\end{equation}
where
\begin{equation}
{\cal D}(\mu X^+)=8\left({\hat \Delta}_m-2\Delta_m\right) \ .
  \label{definedD}
\end{equation}
Recall that ${\hat \Delta}_m$ and $\Delta_m$ tend to $1/12$ and
$-1/48$, respectively, when $m=\mu X^+$ tends to zero. The quantity
${\cal D}(\mu X^+)$ decreases from unity and asymptotes to zero as
$\mu X^+$ increases. Of course, when $\mu$ (and hence $m$) vanishes,
this gives the expected result:
\begin{equation}
  z^2-iX^+X^-=2\pi^2\alpha^\prime\equiv X_H^2 \ .
\end{equation}
Note here that the unusual factor of $-i$ in this expression is as a
result of the Wick rotation, which results in the (complexified)
metric
\begin{equation}
  ds^2=-2idx^+dx^-+\mu^2x^2(dx^+)^2+\sum_{i=1}^{8}dx^idx^i\ .
\label{complexmetric}
\end{equation}
For non--zero $\mu$ it is hard to interpret the result cleanly, but
there is certainly a non--trivial dependence of the location of the
``halo'' on $\mu$, in contrast to the Lorentzian case.

\begin{figure}[htbp]
  \centering
  \includegraphics[height=6cm]{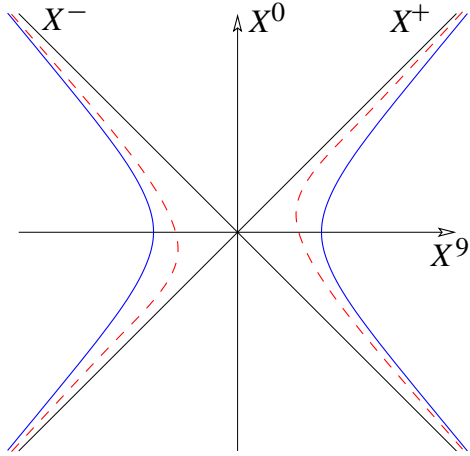}
  \caption{\small 
    The hyperbola (solid curve) represents the edge of the ``halo''
    for D--instantons in flat space, $\mu=0$. For the $\mu\neq0$ case,
    it is deformed to the dashed curve.  The field theory result is
    the pair of lines $X^+X^-=0$. }
  \label{hyperbole}
\end{figure}

As a simple special case, one can place the branes at the same
transverse position, and hence $z^i=0$. Then we have the equation:
\begin{equation}
-iX^+X^-=16\pi^2\alpha^\prime\left({\hat\Delta}_m-2\Delta_m\right)\ .
  \label{simplercase}
\end{equation}
For orientation, let us consider the flat space case $\mu=0$. We can
continue to a more familiar Lorentzian picture by choosing $X^-\to
iX^-$. This gives  a hyperbola in the plane, with equation
\begin{equation}
  \label{circle}
  X^+X^-=2\pi^2\alpha^\prime\equiv X_H^2\ .
\end{equation}
Contrast this to the case of field theory, where the right hand side
would be zero, giving us the light--cone. This is as expected for point
like behaviour. The flat space string theory result gives us a
hyperbola. This is the manifestation of the halo which broadens out
the available region of contact by widening the light--cone into a sort
of ``light--funnel''. For the $\mu\neq0$ case, the hyperbola is
deformed, since $X^-$ decreases more rapidly with increasing $X^+$
than before due to the behaviour of the function ${\cal D}(X^+)$
discussed below equation~\reef{definedD}.  See figure~\ref{hyperbole}.


For the interpretation of the shape of the halo for non--zero $\mu$
once the transverse positions of the branes are different from each
other, more work is needed. This is because the metric is no longer
flat, and furthermore, one has to take seriously the matter of the
Euclidean continuation of the metric implied in the computation of the
amplitude. The choices made mean that the metric is no longer real
(see equation~\reef{complexmetric}), and this presents difficulties of
interpretation which must be explored further.

\section{Discussion}

We have found that the structure of the halo for Lorentzian branes in
the plane wave background is independent of $\mu$, giving the same
physics as for D--branes in flat space. This is because the mass
parameter induced by non--zero $\mu$ in the effective world--volume
theory vanishes in the open string IR limit, the regime where the halo
is to be found. We observed that this is not the case for the
D--instanton (and presumably all Euclidean branes), since their being
pointlike in the $X^\pm$ directions requires the relevant amplitudes
to be defined by starting with the closed string light cone gauge and
then arriving at the open string physics by duality. The resulting
open string physics sees a mass parameter which does not vanish in the
IR limit, and hence the physics of the halo is not the same as in flat
space.  The significance of this non--trivial $\mu$ dependence of the
structure of the halo of the D--instanton (and by extension, all
Euclidean branes defined by starting with the closed string amplitude)
is not clear to us at present. However, it may have some significance,
since D--instantons contribute to type~IIB string theory processes
non--perturbatively
(see {\it e.g.,} ref.\cite{Green:1997tv}). 

\bigskip \bigskip

\centerline{\bf Note on earlier version of this manuscript}
\bigskip

In an earlier version of this manuscript, we noted that there were
$\mu$--dependent effects for Lorentzian branes as well. That was a
mistake, and we apologise for any confusion caused. We misinterpreted
the structure of the amplitudes in
refs.\cite{Bergman:2002hv,Gaberdiel:2002hh}, and treated the effective
mass parameter,  $m$,  in the open string sector as a fixed parameter in
both the Lorentzian and Euclidean cases. This led us to that erroneous
conclusion.

\bigskip

We note that Oren Bergman, Shinji Hirano and Dan Brace, and
additionally Sakura Schafer-Nameki and Aninda Sinha informed us that
they independently observed that our conclusions in the previous
version were faulty and we thank them for comments and discussions.

\section*{Acknowledgements}We thank Matthias Gaberdiel for a conversation.
C.V.J. would like to thank the EPSRC and the PPARC for financial
support. H.G.S. was supported by a doctoral student fellowship from
the Research Council of Norway, by an ORS award, and by the University
of Durham. This paper is report number DCPT-03/13.



\end{document}